\documentstyle[12pt,aaspp4]{article}
\def\beq{\begin{equation}}
\def\eeq{\end{equation}}
\def\ref{\reference}
\def\simge{\mathrel{%
   \rlap{\raise 0.511ex \hbox{$>$}}{\lower 0.511ex \hbox{$\sim$}}}}
\def\simle{\mathrel{
   \rlap{\raise 0.511ex \hbox{$<$}}{\lower 0.511ex \hbox{$\sim$}}}}
\begin{document}
\title{On Radio and X-ray Emission Mechanisms in Nearby, X-ray Bright Galactic Nuclei}
\author{Insu Yi$^{1,2,3}$ and Stephen P. Boughn$^4$}
\affil{$^1$Institute for Advanced Study, Olden Lane, Princeton, 
NJ 08540; yi@sns.ias.edu}
\affil{$^2$Center for High Energy Astrophysics and Isotope Studies, Research
Institute for Basic Sciences}
\affil{$^3$Dept. of Physics, Ewha University, Seoul, Korea; yi@astro.ewha.ac.kr}
\affil{$^4$Department of Astronomy, Haverford College, Haverford,
PA 19041; sboughn@haverford.edu}

\begin{abstract}

It has been suggested that advection-dominated accretion flows (ADAFs) are 
responsible for the X-ray activity in nearby galactic nuclei.  These X-ray
bright galactic nuclei (XBGN) are a heterogeneous group which includes
LINERs, low to moderate luminosity Seyferts, and narrow-line X-ray galaxies with
$2-10$ keV X-ray luminosities in the range $\sim 10^{39}$ to 
$\sim 10^{43}$ erg/s.  In the absence of a radio jet, the core
$15$ GHz radio luminosity of an ADAF is relatively low and roughly proportional 
to the mass of the central black hole.  The predicted radio luminosity
depends primarily on the black hole mass and for XBGN typically falls in the 
range $10^{35}\simle L_R\simle 10^{39}$ erg/s. 
We designate these as ``radio quiet''
XBGN.  However, some level of jet activity seems to be present in most sources 
and the radio emission can be considerably larger than that from the ADAF core. 
We discuss connections between radio-bright XBGN and 
radio-loud, powerful active galactic nuclei (AGN) and suggest
that the radio activities are directly correlated with black hole spins in
both cases.  Even in the presence of a radio jet, high resolution,
high frequency radio imaging of nearby XBGN could identify compact, inverted 
spectrum ADAF radio sources. The unique radio/X-ray luminosity relation is
confirmed in a few cases where black hole masses are known and could be used 
as a tool to estimate unknown black hole masses.  For radio-dim 
($L_R \simle 10^{39}$ erg/s), X-ray bright ($L_x \simge 10^{43}$ erg/s) 
sources, which are primarily Seyferts, the X-ray emission mechanism is not 
accounted for by pure ADAFs and radio activities are likely to be similar 
to those of the radio-quiet AGN.

\end{abstract}


\section{Introduction}

It is widely believed that non-stellar emission in galactic nuclei
indicates the existence of accreting, massive black holes (e.g. Frank et al.
1992). It is, however, unclear how to understand the various emission spectra from 
diverse types of AGN powered by accreting black holes
(e.g. Osterbrock 1989).  Within the luminous AGN population, 
radio luminosities differ greatly and hence the
classification into the radio-loud and radio-quiet sub-populations. 
The luminous optical/UV/X-ray emission is attributed to accretion flows
around massive black holes while the strong radio emission clearly arises from 
powerful radio jets. 

Recently, it has been suggested that the X-ray and radio emission from less 
luminous X-ray bright galactic nuclei (XBGN) such as LINERs and low luminosity
Seyferts could be due to optically thin advection-dominated accretion flows 
(ADAFs) (Yi \& Boughn 1998, Di Matteo \& Fabian 1997, Fabian \& Rees 1995
and references therein). These sources have X-ray luminosities in the range 
$10^{39} \simle L_x \simle 10^{43}$ erg/s. In ADAFs, low-level radio emission
arises from an X-ray emitting, optically thin plasma via 
synchrotron emission (e.g. Narayan \& Yi 1995b).  These sources are
characterized by inverted radio spectra and compact emission regions together 
with hard X-ray spectra (Yi \& Boughn 1998).  Since X-ray and radio 
emission occurs in the same plasma, the X-ray and radio luminosities are
correlated and can be used to estimate the mass of the the central black hole
(Yi \& Boughn 1998).

Radio jets are observed to be widespread in early type radio galaxies (e.g.
Rees et al. 1982).  On the other hand, ADAFs have been shown to have
positive net energy near the rotational axis (Narayan \& Yi 1995a) and,
therefore, are particularly susceptible to outflows.  
If X-ray emission from XBGN is due to ADAFs,
it is plausible that radio jets would also be present and could
dominate the total radio luminosities of these sources.  In fact, Falcke et al.
(1998) recently argued that a large fraction of relatively radio-dim 
active galaxies, QSOs, and LINERs possess parsec scale radio jets. If so, it 
might be the case that the radio luminosities of nearly all XBGN are dominated 
by jet emission.  Never-the-less, high angular resolution observations could 
still reveal the presence of an ADAF radio core with its characteristic 
inverted spectrum.  Many of the sources (XBGN and Seyferts) listed below
are jet dominated sources and, yet, have 15 GHz core luminosities 
similar to those predicted by the ADAF model.  

Franceschini et al. (1998) reported an intriguing correlation between the total 
radio luminosities and the dynamically determined black hole masses
of a small sample of XBGN consisting
primarily of early type galaxies.  They interpreted the correlation
as due to ADAF radio emission around massive black holes accreting from
the hot gas readily available in early type galactic nuclei. Although their
explanation (assuming the accretion rate is determined by the
Bondi rate and the gas density is directly related to the black hole mass) 
is largely implausible (see \S3), 
the correlation does suggest an interesting trend in radio 
properties. Many of their sources are likely to contain radio jets.
By combining them with those used in Yi and Boughn (1998), we attempt to 
distinguish radio-jet from pure ADAF emission in XBGN. 
We examine possible radio/X-ray luminosity relations in pure ADAF flows and in
radio-jet sources with known black hole masses.
Particular attention is paid to the origin of radio activity
and we suggest that among XBGN it is useful to designated two populations, 
radio-bright and radio-dim, which are analogous to more powerful AGN 
populations.

We designate as ``radio-dim'' those sources with 5 GHz luminosities 
$\nu L_{\nu}\simle 10^{38}$ erg/s.  Even if jets are present in these sources 
they are likely to be relatively weak, parsec scale jets and may
still appear as core dominated sources when observed with moderate angular
resolution.  The jet emission in such sources has a flat or inverted spectrum
(e.g. Falck, Wilson, \& Ho 1998).
XBGN with luminosities $L_R\simge 10^{38}$ erg/s are
designated as ``radio-bright'' (to distinguish them from the conventional
``radio-loud'' sources which are much more powerful).
Such sources have more substantial jets and 
will, therefore, invariably appear extended as well as exhibit steeper spectra.

\section{Radio/X-ray Emission from Accreting Massive Black Holes}

{\it Emission from Optically Thin ADAFs:}
In high temperature, optically thin ADAFs, the hard X-ray emission results
from bremsstrahlung and Comptonization (Narayan \& Yi 1995b, Rees et al.
1982). The Compton-upscattered soft photons 
are generated by synchrotron emission which is subject to self-absorption. 
Assuming an equipartition strength magnetic field, all the relevant emission 
components from radio to X-ray are explicitly calculable in terms of black hole
mass, $M_{bh}$, mass accretion rate, ${\dot M}$, and the viscosity parameter,
$\alpha$, for which we adopt the value $0.3$ (Frank et al. 1992).

In ADAFs, radio emission arises directly from synchrotron emission with 
magnetic field
$B\sim 1.1\times 10^4 m_7^{-1/2}{\dot m}_{-3}^{1/2}r^{-5/4}$ Gauss where 
$m=M_{bh}/M_{\odot}$, $m_7=m/10^7$, 
${\dot m}={\dot M}/{\dot M}_{Edd}$, ${\dot m}_{-3}={\dot m}/10^{-3}$, and
$r=R/R_S$ with ${\dot M}_{Edd}=1.39\times 10^{25}m_7$ g/s,
and $R_S=2.95\times 10^{12}m_7~cm$. The optically thin synchrotron emission is 
self-absorbed up to a frequency
$\nu_{syn}(r)\sim 9\times 10^{11}m_7^{-1/2}{\dot m}_{-3}^{1/2}T_{e9}^2
r^{-5/4}Hz$
where $T_{e9}=T_e/10^9K$ is the electron temperature. At radio
frequencies, the luminosity is given by (e.g. Yi \& Boughn 1998 and
references therein)
\beq
L_{R,adv}(\nu)=\nu L_{\nu}^{syn}\sim 2\times 10^{32} x_{M3}^{8/5}m_7^{6/5}
{\dot m}_{-3}^{4/5}T_{e9}^{21/5}\nu_{10}^{7/5}~erg/s
\eeq
where $\nu_{10}=\nu/10^{10}Hz$ and $x_{M3}=x_M/10^3$ is the dimensionless
synchrotron self-absorption coefficient (Narayan \& Yi 1995b). 
Since $x_{M3}\propto (m{\dot m})^{1/4}$ and $T_{e9}\simle 10$ is only weakly
dependent on $m$ \& ${\dot m}$ (e.g. Mahadevan 1996),
we obtain the approximate relation (for $\nu = 15 GHz$)
\beq
L_{R,adv}\sim 3\times 10^{36} m_7^{8/5} {\dot m}_{-3}^{6/5}~erg/s.
\eeq

The 2-10 keV X-ray emission from ADAFs is due to bremsstrahlung and
Comptonization of synchrotron photons. At low mass accretion rates,
${\dot m}\simle 10^{-3}$, the X-ray luminosity has a significant 
bremsstrahlung contribution whereas at relatively high mass accretion rates
$10^{-3}\simle {\dot m}\simle 10^{-1.6}$, Comptonization dominates in the
2-10 keV band. ADAFs can only exist for mass accretion rates below a critical
value, ${\dot m} < {\dot m_{crit}} \approx 10^{-1.6}$ (Rees et al. 1982;
Narayan \& Yi 1995).  Yi \& Boughn (1998) have shown that the 2-10 keV X-ray 
luminosity is related to the 15 GHz radio luminosity by a simple relation
\beq
L_{R,adv}\sim 1\times 10^{36} m_7 (L_{x,adv}/10^{40} erg~s^{-1})^{y}
\eeq
where $y\sim 1/5$ for systems with ${\dot m}<10^{-3}$ and $y\sim 1/10$ for 
systems with ${\dot m}>10^{-3}$. For our discussions, we adopt $y=1/7$
which is a reasonably good approximation 
for $10^{-4}\simle {\dot m}\simle 10^{-1.6}$.
The bolometric luminosity, which is dominated
by the X-ray luminosity for ${\dot m}\simge 10^{-3}$, is roughly given by
$L_{adv}\sim 30{\dot m}^2L_{Edd}$ where $L_{Edd}=0.1{\dot M}_{Edd}c^2$
(e.g. Yi 1996).

{\it Emission from Optically Thick Disks:}
In ADAFs, optical/UV emission is characteristically weak which distinguishes
ADAFs from the high radiative efficiency accretion disks commonly assumed 
for luminous AGN. For the high accretion rates required by luminous AGN, 
i.e. ${\dot m}\simge 10^{-1.6}$, ADAFs do not
exist (Narayan \& Yi 1995b, Rees et al. 1982). 
It is widely assumed that at such high rates,
accretion takes the form of a geometrically thin, optically thick
accretion flow with a hot, X-ray emitting corona (Frank et al. 1992).  Then
\beq
L_{x,disk}\sim \eta_{eff}{\dot M}c^2\sim 1.3\times 10^{42}(\eta_{eff}/0.1)m_7
{\dot m}_{-3}~erg/s
\eeq
where $\eta_{eff}$ is the radiative efficiency of the accretion flow.
The efficiency must be high, $\eta_{eff}\sim 0.1$, to account for the 
observed X-ray luminosities.

{\it Radio Jet Power:}
Most AGN have radio luminosities that far exceed those predicted by the ADAF 
model (see, for example, Fig. 1).  That there exists a wide range of radio 
luminosities for a relatively narrow X-ray luminosity range (of X-ray
selected sources) is likely the 
result of radio jets of various strengths; however, it is still unclear just
how radio-emitting jets are powered.

Given the fact that ADAFs are prone to outflows/jets (Narayan \& Yi 1995a,
Rees et al. 1982), it is likely that many ADAF sources have radio-jets.
Neither the ADAF nor the thin disk models can self-consistently account for 
this radio emission.  However, if the radio-jet is 
powered by a rotating black hole accreting from a magnetized plasma, as is 
generally believed, the radio power can be described by the Blandford-Znajek 
process (e.g. Frank et al. 1992), according to which
$L_{R,jet}\propto {\bar a}^2 M_{bh}^2 B^2$ or
\beq
L_{R,jet}\sim 1.2\times 10^{42} {\bar a}^2 \epsilon_{jet} m_7 {\dot m}_{-3}
\eeq
where
${\bar a}\le 1$ is the black hole spin parameter and $\epsilon_{jet}\le 1$ 
is the efficiency of the radio emission. 

\section{Radio/X-ray Luminosity Relation and Black Hole Masses}

Fig. 1 is a plot of the ratio of the 5 GHz total (as opposed to ``core'') radio 
luminosity to 2-10 keV X-ray luminosity vs. X-ray luminosity for 
a collection of LINERs, moderate to low 
luminosity Seyferts, X-ray bright elliptical galaxies, and the weak nuclear
sources Sgr A$^*$ and M31.  The sources were compiled from Yi \& Boughn (1998) 
and Franceschini et al. (1998).  X-ray fluxes were converted to 2-10 keV fluxes 
in those cases where the data were in a different band and are uncertain by a 
factor of a few at most. The 5 GHz radio fluxes are from the Green Bank survey
(Becker et al. 1991; Gregory \& Condon 1991).  The solid lines 
are predicted for ADAFs of different black hole 
masses and are discussed in more detail below and in Yi \& Boughn (1998). 

Dynamical black hole mass estimates are available for NGC1068, 1316, 4258, 
4261, 4374, 4486, 4594, M31, and Sgr A$^*$.  The mass of NGC1068 is from
Greenhill et al. (1996); of NGC4258 from Herrnstein et al. (1998); of
Sgr A$^*$ from Eckart \& Genzel (1997); of M31, M87, and NGC4594 from
Richstone et al. (1998); of NGC 4261 from
Richstone et al. (1998) and Ferrarese, Ford, \& Jaffe (1996); and
of NGC1316 and NGC4374 from Franceschini, Vercellone, \& Fabian (1998). 
Uncertainties in these masses are not easily quantified; however, from the 
spread of different mass estimates it seems likely that they are accurate to
within a factor of 2.
Fig. 2 depicts the correlation of 5 GHz radio luminosity and black hole 
mass that was noted by Franceschini et al. (1998).  They also noted
that the correlation of X-ray luminosity and black hole mass is very weak
(see Fig. 3).

Franceschini et al. (1998) argue that ADAF radio emission is responsible for 
the correlation between radio luminosity and black hole mass. They assume that 
the density of the accreted matter at large distances from the black hole is
proportional to the black hole mass, i.e. $\rho_{\infty}\propto M_{bh}$, 
and that $M_{bh}\propto M_{gal}\propto
c_{s,\infty}^4$ where $c_{s,\infty}$ is the sound speed of the accreting
gas. The latter relation is adopted from the Faber-Jackson relation.
Assuming Bondi accretion, then ${\dot M}\propto M_{bh}^2\rho_{\infty}/c_s^3
\propto M_{bh}^{9/4}$.  If one ignores the dependence of $x_{M3}$ on ${\dot M}$,
eq. 1 implies that $L_{R,adv}\propto M_{bh}^{11/5}$.  This is the
power-law mass-luminosity relation derived by Francheschini et al. (1998) and 
corresponds to the dashed line in Fig. 2.  The power law appears to agree with
the trend in the data although its statistical significance is obviously
limited due to the sample size and uncertainties in the radio fluxes (If one 
includes the dependence of $x_{M3}$ on ${\dot M}$ then the power law slope is
changed somewhat but the following conclusions are the same).

However, such an explanation inevitably predicts a strong correlation between
$L_x$ and $M_{bh}$.  For ADAFs $L_{x,adv}\propto m{\dot m}^{x}$ 
where $x=2$ if X-rays come from bremsstrahlung and $x>2$ if X-rays are 
from multiple Compton scattering (Yi \& Boughn 1998; Yi 1996).  Therefore, 
$L_{x,adv}\propto M_{bh}^{(5x+4)/4}$ and
$L_{R,adv}/L_{x,adv}\propto L_{x,adv}^{(24-25x)/(20+25x)}$.
For a wide range of $x$, $2 \le x \le 10$, we expect
$L_{R,adv}/L_{x,adv} \propto L_{x,adv}^{-0.4}$ to $L_{x,adv}^{-0.8}$.
Therefore, all sources shown in Franceschini et al. (1998) should
fall in a single band with slope of $\sim -0.6$ in $L_R/L_x$ vs. $L_x$ plane.
This is not evident in Fig. 1.  The predicted X-ray mass-luminosity relation
is quite steep, $L_{x,adv} \propto M_{bh}^{\beta}$, with 
$3.5 \le \beta \le 13.5$ for $2 \le x \le 10$ and is not consistent with 
observations as indicated in Fig. 3 where the dashed line 
is for $x=4$, i.e. $\beta = 6$.  Apparently, the most massive black 
hole sources are too X-ray dim to be compatible with the Francheschini et al. 
(1998) model.  If the measured radio luminosities are indeed from ADAFs, the 
observed black hole masses predict much higher $L_x$ than observed.
Therefore, the correlation found by Franceschini et al. (1998) cannot
be attributed to ADAFs powered by Bondi accretion. 
An alternative explanation is that the observed radio luminosities are
due to much more energetic sources, e.g. jets, whose luminosities are not 
directly related to the radio luminosity of the ADAF core.

If the excess radio emission is due to jet activity, then eq. 5 and
$L_{x,adv}\propto m{\dot m}^{x}$ imply
\beq
L_{R,jet}\propto {\bar a}^2L_{x,adv}^{1/x}M_{bh}^{(x-1)/x}
\eeq
and for a wide range of $x\ge 4$ we expect the dominant scaling 
$L_{R,jet}\propto {\bar a}^2M_{bh}$. Then the observed $L_R$ vs. $M_{bh}$ 
plot in Fig. 2 could simply be a result of the combination of 
$L_{R}\propto M_{bh}$ and a distribution of ${\bar a}$'s (dotted lines).
If the accretion rate is controlled 
by the Bondi rate, i.e. ${\dot M}\propto M_{bh}^{9/4}$, then
$L_x\propto M_{bh}^{(5x+4)/4}$ and $L_{R,jet}\propto {\bar a}^2M_{bh}
^{9/4}$ which is similar to the $L_{R,adv}\propto M_{bh}^{11/5}$ 
relation of Franceschini et al. (1998).

The radio and X-ray emission from an ADAF are given by
$L_{x,adv}\propto m{\dot m}^{x}$, $L_{R,adv}\propto M_{bh}^{8/5}{\dot m}^{6/5}$
(eq. 2), and, therefore, $L_{R,adv}\propto m^{(8x-6)/5x}L_{x,adv}^{6/5x}$
(Yi \& Boughn 1998).  These trends are shown as solid lines in Figs. 1 and 3. 
In Fig. 3, the X-ray emission for most of
the sources is well accounted for if ${\dot m}$ varies from $10^{-2}$ to 
$10^{-3}$.  Since ADAF radio emission is highly localized ($\ll 1$ pc), it is 
not surprising that the total radio fluxes plotted in Figs. 1 exceed
those predicted by ADAFs.  Figs. 4 and 5 are plots of the 15 GHz ``core'' 
luminosities of these same sources. In a few cases, radio fluxes were converted 
from 5 GHz to 15 GHz using the $\nu^{7/5}$ power-law of eq. 1. While this
conversion is inappropriate for steep spectrum sources, it allows a direct 
comparison of these luminosities with those predicted for an ADAF. In any case,
such values are in error by at most a factor of a few 
(If a source has a steep spectrum, i.e. $\nu L_{\nu} \propto \nu^{0.2}$, the
error is less than a factor of 4).
The solid lines in Figs. 4 and 5 are the same as those in Figs. 3 and 1 
and they are, indeed, compatible with most of the sources; however, only in 
three cases is the angular resolution good enough to approximately resolve the 
ADAF core.  The sources are discussed individually in \S4.

The ratio of jet to ADAF radio emission depends only weakly on black hole mass
and accretion rate.  From eqs. 2 and 5,
\beq
L_{R,jet}/L_{R,adv}\sim 4\times 10^5 {\bar a}^2 \epsilon_{jet}
m_7^{-1/5}{\dot m}_{-3}^{1/5}
\eeq
i.e. $L_{R,jet}$ can far exceed $L_{R,adv}$ for sources with typical $M_{bh}$ 
and ${\dot M}$ unless ${\bar a}\ll 2\times 10^{-3} \epsilon_{jet}^{-1/2}$.
This appears to be the case for many of the sources in Fig. 1.

If $L_x$ is from a thin disk and $L_R$ is from a jet (since a thin disk itself 
is unable to emit in the radio) we expect (from eqs. 4 and 5)
\beq
L_{R,jet}/L_{x,disk}\sim {\bar a}^2 \epsilon_{jet}/\eta_{eff}\le \epsilon_{jet}
/\eta_{eff}.
\eeq
For radio-bright galaxies with ${\bar a}\sim 1$, this ratio is independent of 
X-ray luminosity or black hole mass. For radio-dim sources 
(e.g. ${\bar a}\ll 1$), we expect $L_x\gg L_R$.

\section{Discussion of Individual Sources}

{\it M87, NGC4258, and Sgr A$^*$}

These three sources have estimated black hole masses and, in
addition, have been the subject of high angular resolution radio observations
(VLBI, VLBA, and VLA respectively).  While M87 and NGC4258 have relatively
strong, extended radio emission from jets, the high spatial resolution 
($\ll$ 1 pc)
of the radio observations affords a nearly resolved view of the hypothetical
core ADAF.  The masses implied by their locations on 
Fig. 4 are within a factor of two of the dynamical
estimates for these two sources.  

All three of these sources have been previously identified as ADAF
candidates (Reynolds et al. 1997, Lasota et al. 1996, and Narayan et al. 1995).
However, Herrnstein et al. (1998) have argued that the NGC4258 core source used 
in this paper is actually one of two, unresolved radio jets located 
$\sim 0.01$pc from the warped plane of the accretion disk.  If future 
observations strengthen this conclusion then the ADAF mechanism for
NGC4258 will be called into question (cf. Blackman 1998).

Sgr A$^*$ is distinct from the other sources discussed in this paper in that
it has a much smaller X-ray luminosity due presumably to a low accretion
rate.  Never-the-less, Narayan, Yi, $\&$ Mahadevan (1995) have modeled it as an
ADAF so we include it here as an example of a low mass, low accretion rate
ADAF.  We have plotted in Figs. 1, 3, and 4 the ROSAT X-ray flux corrected for 
extinction (Prehehl \& Trumper 1994).  The uncertainty in the extinction 
corrected X-ray flux has little consequence since for a given core radio flux, 
the X-ray flux is nearly independent of black hole
mass.  The location of the Sgr A$^*$ point would simply be translated
along a line that is nearly parallel to the theoretical curves in Figs. 1 and 4.

{\it NGC1068, 1316, 4261, 4374, 4594, M31}

These are the remaining sources for which there are dynamical estimates of
the masses of the central black holes; although, none has been observed in
the radio with very high angular resolution.  NGC4594 (M104) is a radio-dim, 
flat spectrum source with no sign of jet activity.  VLA observations (Hummel 
et al. 1984) indicate a core dominated source with a size $<1$ pc.  If this 
radio source is interpreted as an ADAF then the black hole mass implied by 
Fig.4 is the same as the dynamical estimate.   NGC1316, 4261, and 4374, 
on the other hand, are radio-bright sources (see Fig. 1).  NGC1316 is a
bright, lobe dominated radio galaxy.  Because of relatively poor angular
resolution, its core flux is not well defined; however, the radio luminosity
within 3 to 5 arcsec is within a factor of two of that predicted for an 
ADAF flow with the measured black hole mass.  NGC4261 and 4374 are also
extended sources but again with core luminosities comparable (within a
factor $\sim 2$) to that predicted by the ADAF model.  Considering the 
uncertainties of the X-ray and radio observations and of the dynamical mass 
estimates, the level of agreement is impressive.

It is clear from Fig. 4 that NGC1068 and M31 do not fit the ADAF model; 
however, both are unusual sources.  The direct X-rays from the nucleus of
NGC1068 appear to be highly absorbed with the observed X-ray flux consisting
entirely of scattered photons.  The inferred intrinsic X-ray luminosity is 
likely to be $L_x\simge 5\times 10^{43} erg/s$ (e.g. Koyama et al. 1989).
The Eddington luminosity for the estimated black hole mass,
$M_{bh}\sim 2\times 10^7M_{\odot}$, is $L_{Edd}\sim 2\times 10^{45} erg/s$; 
therefore,  $L_x\simge 3 \times 10^{-2}L_{Edd}$.  This implies a bolometric
luminosity in excess of the maximum allowed for ADAF flows (Narayan, 
Mahadevan \& Quataert 1998) so it is unlikely that the 
X-ray emission is from an ADAF.  NGC1068 has a resolved central core with an 
observed 15 GHz luminosity of $\sim 7\times 10^{37} erg/s$ (Sadler et al. 1995)
together with clear jet structures.  If the X-ray 
emission occurs with high efficiency $\sim 10$\%, the observed $L_x$ implies 
${\dot m}\sim 2\times 10^{-2}$. In this case, 
$L_{R,jet}\le 5\times 10^{43}{\bar a}^2\epsilon_{jet}$ which is far more than 
the observed $L_R$ even with $\epsilon_{jet}\ll 1$ for ${\bar a}\sim 1$. 
Therefore, NGC1068 is likely to be a typical radio-quiet, luminous Seyfert
(Falcke et al. 1998).

M31 has an extremely low core radio luminosity, $5 \times 10^{32} erg/s$
(Gregory \& Condon 1991).  If this is attributed to an ADAF with
$M_{bh}=3\times 10^7M_{\odot}$, then the implied accretion rate is very small,
${\dot m}\sim 10^{-6}$, and the expected ADAF X-ray luminosity is 
$L_x\sim 4\times 10^{34} erg/s$, much less than that observed.  In this 
case it is possible that the observed $L_x$ is dominated by a few bright X-ray 
binaries similar to those recently reported in M32 (Loewenstein et al. 1998).
Indeed, Sgr A$^*$ is surrounded by bright, resolved X-ray sources 
(e.g. Genzel et al. 1994) and if put at the distance of M31 would appear as a
brighter, unresolved core.  Finally, it is not at all clear whether
the two temperature ADAF model is valid at such low accretion rates.

{\it NGC3031, 3079, 3627, 3628, 4736, and 5194}

These sources are all nearby LINERs with similar 2-10 keV X-ray luminosities,
$6\times 10^{39} < L_x < 3\times 10^{40}$.  There are no dynamical mass
estimates for the black holes in any of them.  It is interesting to
note, however, that with the exception of NGC3079 the core radio luminosities
are consistent with those predicted by ADAFs for black hole masses of
$\sim 10^7$ to $\sim 10^8$ $M_{\odot}$ (see Fig. 4). 
The larger luminosity of NGC3079 
would require $M_{bh} = 1.4 \times 10^9 M_{\odot}$.  In addition, all of the
sources have either flat or inverted radio spectra.

None of these is a particularly strong radio source; however, only one of them,
NGC3031, is dominated by core emission. 
Coincidentally, this is the only source 
for which there are high angular resolution (VLBI) radio
observations (Bietenholz et al. 1996; Reuter \& Lesch 1996; Turner \& Ho 1994). 
The spectrum is
inverted ($\propto \nu ^{1/3}$) up to 100 GHz where it begins to turn over.  
This is consistent with an ADAF.  At 22 GHz the core is barely resolved, 
$\sim 0.1 mas$, which implies a linear size of $\sim 0.002 pc$.  This is also
consistent with an ADAF.  If the radio and X-ray luminosities are, indeed, due
to an ADAF then the implied black hole mass is $\sim 1 \times 10^8 M_{\odot}$.
It will be interesting to see if future dynamical analyses confirm this value.

{\it NGC3227, 4151, 5548, and 4388}

These four sources are all classified as Seyferts and none is a 
particularly strong radio source (see Fig. 1).  Their radio and X-ray 
luminosities are consistent with ADAFs onto $(1-5)\times 10^8 M_{\odot}$ 
black holes and with accretion rates below the critical value (Fig. 4).  
Because the ADAF cores are unresolved, it is certainly possible
that the radio core luminosities and hence masses are both overestimates.
However, it seems unlikely that such corrections would result in moving these 
sources from the ADAF region.

These three sources are on average
$\sim 600$ times more X-ray luminous than the six LINERs discussed above 
but have only twice the core radio luminosity (see Fig. 4).  
There are no black hole mass estimates for any of these sources so it is
not possible to make quantitative comparisons of their emissions with those
predicted by the ADAF model.  However, to the extent that the average 
black hole masses of these sources are comparable we note that the narrow 
range of core radio luminosity is qualitatively consistent with the weak 
dependence of $L_R$ on $L_x$ (see eq. 3).  The wide range in X-ray luminosities 
is then presumably due to differences in the accretion rates for these sources.

It should be emphasized that there is considerable uncertainty in the above 
comparisons of dynamical mass estimates with ADAF predictions.  In addition
to the observational and modeling uncertainties, one must contend with the
intrinsic variability of ADAF sources (Blackman 1998, Ptak et al. 1998).
Multiple epoch radio and X-ray observations are
necessary to quantify the extent of the variability and to estimate the mean
fluxes. For these reasons, the good agreement between the mass estimates and
ADAF predictions in this paper may be partly fortuitous.

\section{Classification of X-ray Bright Galactic Nuclei}

The previous discussion suggests a useful classification scheme based on the 
radio and X-ray luminosities of XBGN and moderate luminosity Seyferts.  From 
Figs. 1 and 6 it is clear that the 5 GHz ADAF (total) radio luminosity
$L_R\simle 10^{38}erg~s^{-1}$ 
unless $M_{bh}\simge 10^9 M_{\odot}$ and the accretion
rate is near the maximum allowed by ADAFs.  Therefore, we designate a source as 
``radio-bright'' if its total (core plus jet) 5 GHz radio luminosity satisfies 
$L_R > 10^{38}erg s^{-1}$.  Note that this doesn't preclude XBGN with
luminosities below this value from being jet dominated; in fact, of the sources
in Fig. 1 only NGC3031 and 4594 are ADAF core dominated.  However, XBGN 
designated as radio-bright will certainly be jet dominated radio sources.  
Fig. 6 illustrates this classification for the same sources as in Fig. 1.
The upper dashed curve corresponds to the ADAF luminosity for a 
$10^9 M_{\odot}$ black hole (which is motivated by the fact that black holes 
with $M_{bh}>10^9M_{\odot}$ are rare) and the lower dashed curve corresponds 
to maximally accreting ADAFs, i.e. ${\dot m} = 10^{-1.6}$. Sources falling 
within the region at the bottom left of the diagram (bounded by the 
dashed curves) are designated ``radio-dim'' XBGN.  As noted, these are 
presumably ADAFs with little to moderate jet activity.  Sources to the left in 
this region have low accretion rates while sources at the bottom have low mass 
central black holes.  Above this region are the radio-bright XBGN discussed 
above.  Sources at the right of the diagram are too X-ray 
bright to be the result of an ADAF but rather require the high 
efficiency accretion mechanism of powerful AGN.
These classifications are not unambiguous.  For example, an AGN with a low
mass central black hole and low X-ray luminosity but with moderate jet activity
could appear in the upper left part of the ``radio-dim XBGN'' region even 
though the radio luminosity far exceeds that predicted by the ADAF mechanism.
High angular resolution radio observations, an estimate of black hole mass, 
and/or the determination of the radio spectral index would likely distinguish 
between these two cases.  

\section{Conclusions}

For a given black hole mass, the ADAF model predicts a unique radio/X-ray 
luminosity relation.  ADAF radio emission is essentially characterized by an
inverted spectrum and a very compact emission region,
$\ll 1pc$.  So far, the only sources for which there are both high angular 
resolution radio data and dynamical estimates are M87, NGC4258 and Sgr A$^*$.
The observations for all three are consistent with the predictions of the ADAF
emission model.  There are four other sources (NGC1316, 4261, 4374, and 4594)
for which moderate angular resolution radio observation and dynamical black hole
mass estimates are available.  The X-ray and core radio luminosities of these
XBGN are also consistent with the ADAF model.  Considering
observational and modeling uncertainties, the agreement is quite good.

In the future, high angular resolution radio observation combined with X-ray 
observations might enable the central black hole masses of XBGN to be 
estimated.  For example, the nuclear source in NGC3031 (M81) is both small 
($\sim 0.002$ pc) and has an inverted spectrum (Reuter \& Lesch 1996).  
Based on its X-ray and core radio flux, the ADAF implies a black hole mass of 
$M_{bh} \sim 1 \times 10^8 M_{\odot}$ (see Fig. 4).  
Although source variability has not yet
been taken into account, we suggest that future dynamical estimates of the
central black hole of NGC3031 won't be far from this value.

Multiple epoch, high angular resolution, high frequency radio 
observations will be crucial to test the ADAF mechanism and, subsequently, 
to estimate central black hole masses. Even with such 
observations, it will still be difficult to characterize sources if they are 
strongly obscured as expected in Seyfert 2's.  Relatively low X-ray
luminosity sources will be detected by AXAF and will greatly improve the test 
of the ADAF paradigm among LINERs and other relatively low luminosity XBGN 
with $L_x\simle 10^{40} erg/s$.

The compact, inverted spectrum characteristics of ADAFs can also be used to 
distinguish them from small scale jets which are abundant in XBGN.  
A moderate excess of radio emission from such a source probably indicates 
the presence of low level (parsec scale) jet activity.  
We propose an XBGN classification scheme analogous to the 
radio-loud/radio-quiet classification
of powerful AGN.  Whereas, it is likely that the radio luminosities for all 
powerful AGN (radio-quiet and radio-loud) are due to jets, it is possible 
for some radio-dim XBGN to be dominated by an ADAF core.  On the other hand,
radio-bright XBGN are undoubtedly jet dominated.  ADAFs are strong, hard
X-ray sources and to the extent that they are present in XBGN, it is likely 
that the $2-10$ keV core luminosities of XBGN will be dominated by 
ADAF emission.

\acknowledgments
We would like to thank R. van der Marel and Douglas Richstone for useful 
information on black hole masses, Pawan Kumar and Tal Alexander for help 
with NGC1068, and Ramesh Narayan for early discussions on spectral states 
of accreting massive black holes. This work was supported in part by the 
SUAM Foundation (IY) and NASA grant\# NAG 5-3015 (SB).

\clearpage

\noindent

\vfill\eject
\clearpage

\centerline{\bf Figure Captions}

\vskip 0.3cm\noindent
Figure 1: 
The ratio of 5 GHz radio luminosity $L_R$ to $2-10$ keV X-ray luminosity 
$L_x$ vs. $L_x$ for XBGN.
The radio luminosities include both core and extended contributions. Scaling 
from other frequencies is discussed in the text. The sources are adopted from
Yi \& Boughn (1998) and Franceschini et al. (1998). The solid lines are 
the ADAF predictions for black hole masses of $10^6$ to $10^9 M_{\odot}$ 
(marked by log values at the bottom of each curve). For each line, 
the dimensionless mass accretion rate ${\dot m}$ varies from $10^{-4}$ 
(upper dotted line) to $10^{-1.6}$ (lower dotted line), the maximum allowed
ADAF accretion rate (as in Yi \& Boughn 1998).

\vskip 0.3cm\noindent
Figure 2:
5 GHz $L_R$ vs. $M_{bh}$ for the nine sources with black hole mass 
estimates.  The radio fluxes are as in Fig. 1. The dashed line is the 
$L_R\propto M_{bh}^{11/5}$ relation suggested in Franceschini et al. (1998).
The two dotted lines depict the $L_R\propto {\bar a}^2M_{bh}$ relation 
expected for radio-jet sources with a spread of a factor $\sim 10$ in 
${\bar a}^2$ (see text). The references for the dynamical mass estimates are
listed in the text.

\vskip 0.3cm\noindent
Figure 3:
2-10 keV $L_x$ vs. $M_{bh}$ for the same sources as in Fig. 2. The dashed 
line is the steep correlation $L_x\propto M_{bh}^6$ expected if the 
Franceschini et al. (1998) model were valid. The three solid lines 
correspond to $L_x\propto M_{bh}$ correlations for 
${\dot m}\sim 10^{-2}, 10^{-3}, 10^{-4}$ from top to bottom (see text).
The references for the dynamical mass estimates are listed in the text.

\vskip 0.3cm\noindent
Figure 4:
The ratio of 15 GHz core radio luminosity $L_R$ to $2-10$ keV X-ray luminosity 
$L_x$ vs. $L_x$ for the sources in Fig. 1.  The solid and dotted lines
are the same ADAF predictions as in Fig. 1.  Note: ${\dot m} \sim 10^{-1.6}$ 
is the maximum allowed ADAF accretion rate.

\vskip 0.3cm\noindent
Figure 5:
15 GHz core radio luminosity vs. $M_{bh}$ for the nine sources in Fig. 2. 
The three solid lines correspond to the ADAF-based relation 
$L_R\propto {\dot m}^{6/5}M_{bh}^{8/5}$ for the three ${\dot m}$
values shown in Figure 3. The references for the dynamical mass estimates are
listed in the text.

\vskip 0.3cm\noindent
Figure 6:
Schematic classification of XBGN sources according to 5 GHz total radio 
and 2-10 keV X-ray luminosities.  The sources are the same as in Fig. 1.
The upper dashed line corresponds to the the ADAF $L_R$ 
for $M_{bh}=10^9M_{\odot}$ and the lower dashed line corresponds to the
maximum ADAF accretion rate, ${\dot m}=10^{-1.6}$.
The region on the right is occupied by powerful AGN in which the 
ADAF mechanism cannot be operating.


\begin{references}
\reference{A}  Becker, R. H., White, R. L., \& Edward, A. L. 1991, ApJS, 75, 1
\reference{A} Bietenholz et al. 1996, ApJ, 457, 604
\reference{A} Blackman, E. G. 1998, MNRAS, 299, L48
\reference{A} Di Matteo, T. \& Fabian, A. C. 1997, MNRAS, 286, 393
\reference{A} Eckart, A. \& Genzel, R. 1997, MNRAS, 284, 576
\reference{A} Fabian, A. C. \& Rees, M. J. 1995, MNRAS, 277, L55
\reference{A} Falcke, H., Wilson, A. S., \& Ho, L. C. 1998, to appear in the
Proceedings of the Workshop on Relativistic Jets in AGN, eds. M. Ostrowski,
M. Sikora, G. Madejski, M. Begelman (astro-ph/9708126)
\reference{A} Ferrarese, L., Ford, H. C., \& Jaffe, W. 1996, ApJ, 470, 444
\reference{A} Franceschini, A., Vercellone, S., \& Fabian, A. C. 1998, MNRAS, in
press (astro-ph/9801129)
\reference{A} Frank, J., King, A. R., and Raine, D. 1992, Accretion Power 
in Astrophysics, Cambridge: Cambridge University Press
\reference{A} Genzel, R., Hollenbach, D., \& Townes, C. H. 1994, Rep. Prog. 
Phys., 57, 417
\reference{A} Greenhill, L. J., Gwinn, C. R., Antonucci, R., \& Barvainis, R.
1996, ApJ Letters, 472, L21
\reference{A} Gregory, P. D., \& Condon, J. J. 1991, ApJS, 75, 1011
\reference{A} Herrnstein, J. R., Greenhill, L. J., Moran, J. M., Diamond, P. J.,
Inoue, M., Nakai, N., \& Miyoshi, M. 1998, ApJ Letters, in press 
(astro-ph/9802264)
\reference{A} Hummel, E., van der Hulst, J. M., \& Dickey, J. M. 1984, A\& A, 
134, 207
\reference{A} Koyama, K. et al. 1989, PASJ, 41, 731
\reference{A} Koyama, K. et al. 1996, PASJ, 48, 249
\reference{A} Lasota, J.-P., Abramowicz, M. A., Chen, X., Krolik, J., 
Narayan, R., \& Yi, I. 1996, ApJ, 462, 142
\reference{A} Loewenstein, M., Hayashida, K., Toneri, T., \& Davis, D. S. 1998,
ApJ, 497, 681
\reference{A} Mahadevan, R. 1996, ApJ, 465, 327
\reference{A} Narayan, R., Mahadevan, R., \& Quataert, E. 1998, The Theory of
	Black Hole Accretion Disks, eds. M. A. Abramowicz, G. Bjornsson, \&
	J. E. Pringle (Cambridge University Press).
\reference{A} Narayan, R. \& Yi, I. 1995a, ApJ, 444, 231
\reference{A} Narayan, R. \& Yi, I. 1995b, ApJ, 452, 710
\reference{A} Narayan, R., Yi, I., \& Mahadevan, R. 1995, Nature, 374, 623
\reference{A} Osterbrock, D. E. 1989, Astrophysics of Gaseous Nebulae and
Active Galactic Nuclei, Mill Valley: University Science Books
\reference{A} Predehl, P., \& Trumper, J. 1994, A \& A 290, L29
\reference{A} Ptak, A., Yaqoob, T., Mushotzky, R., Serlemitsos, P., \&  
Griffith, R. 1998, ApJL, 501, L37
\reference{A} Rees, M. J., Begelman, M. C., Blandford, R. D., Phinney, E. S.
1982, Nature, 295, 17
\reference{A} Reuter, H.-P., \& Lesch, H. 1996, A\&A, 310, L5
\reference{A} Reynolds, C. S., Di Matteo, T., Fabian, A. C., Hwang, U., \& 
Canizares, C. R. 1996, MNRAS, 283, L111
\reference{A} Sadler, E. M., Slee, O. B., Reynolds, J. E., \& Roy, A. L. 1995,
MNRAS, 276, 1373
\reference{A} Richstone, D. et al. 1998, Nature 395, A14
\reference{A} Turner, J. L., \& Ho, P. J. 1994, ApJ, 421, 122
\reference{A} Yi, I. 1996, ApJ, 473, 645
\reference{A} Yi, I. \& Boughn, S. P. 1998, ApJ, 499, 198
\end{references}
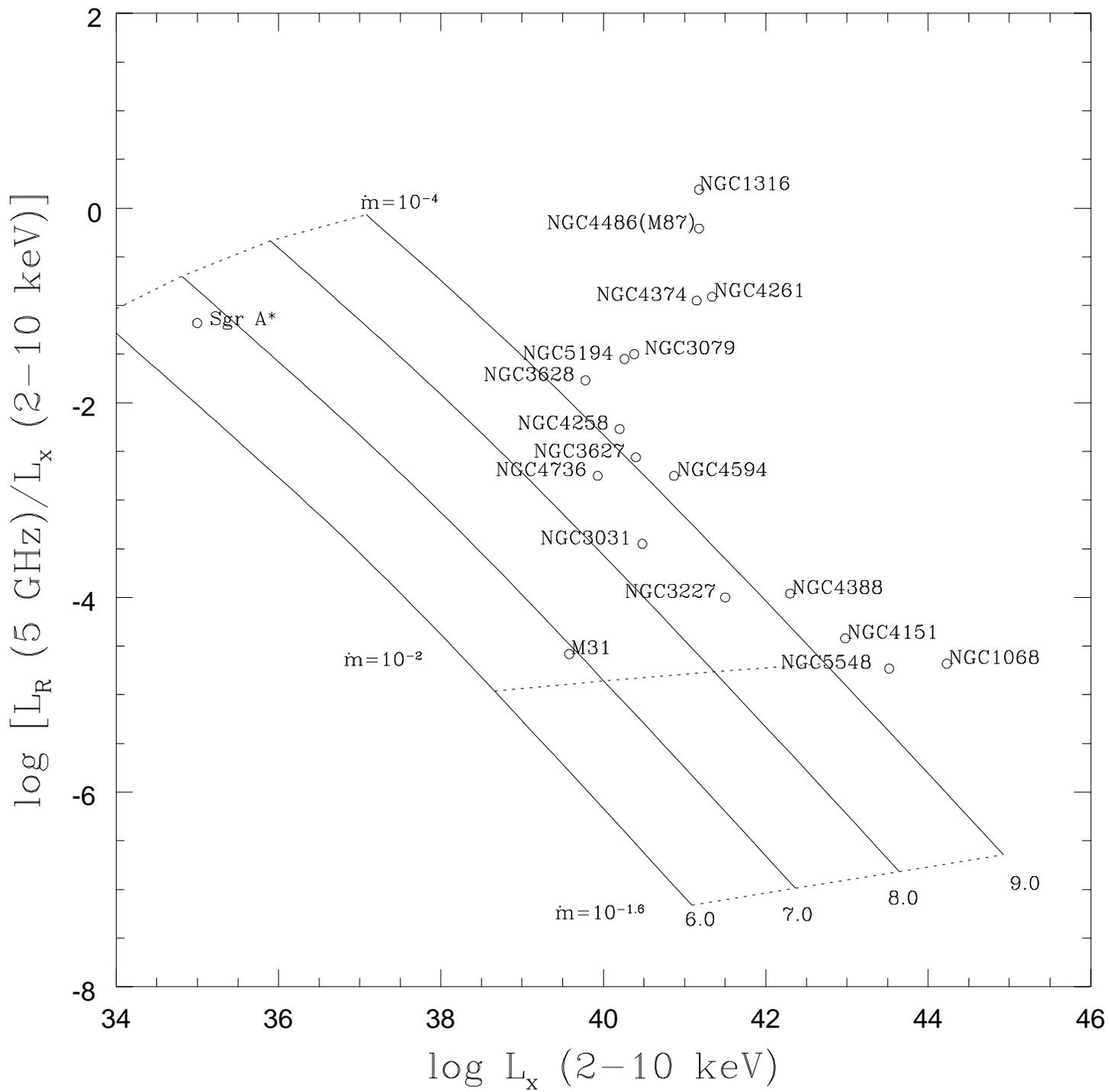
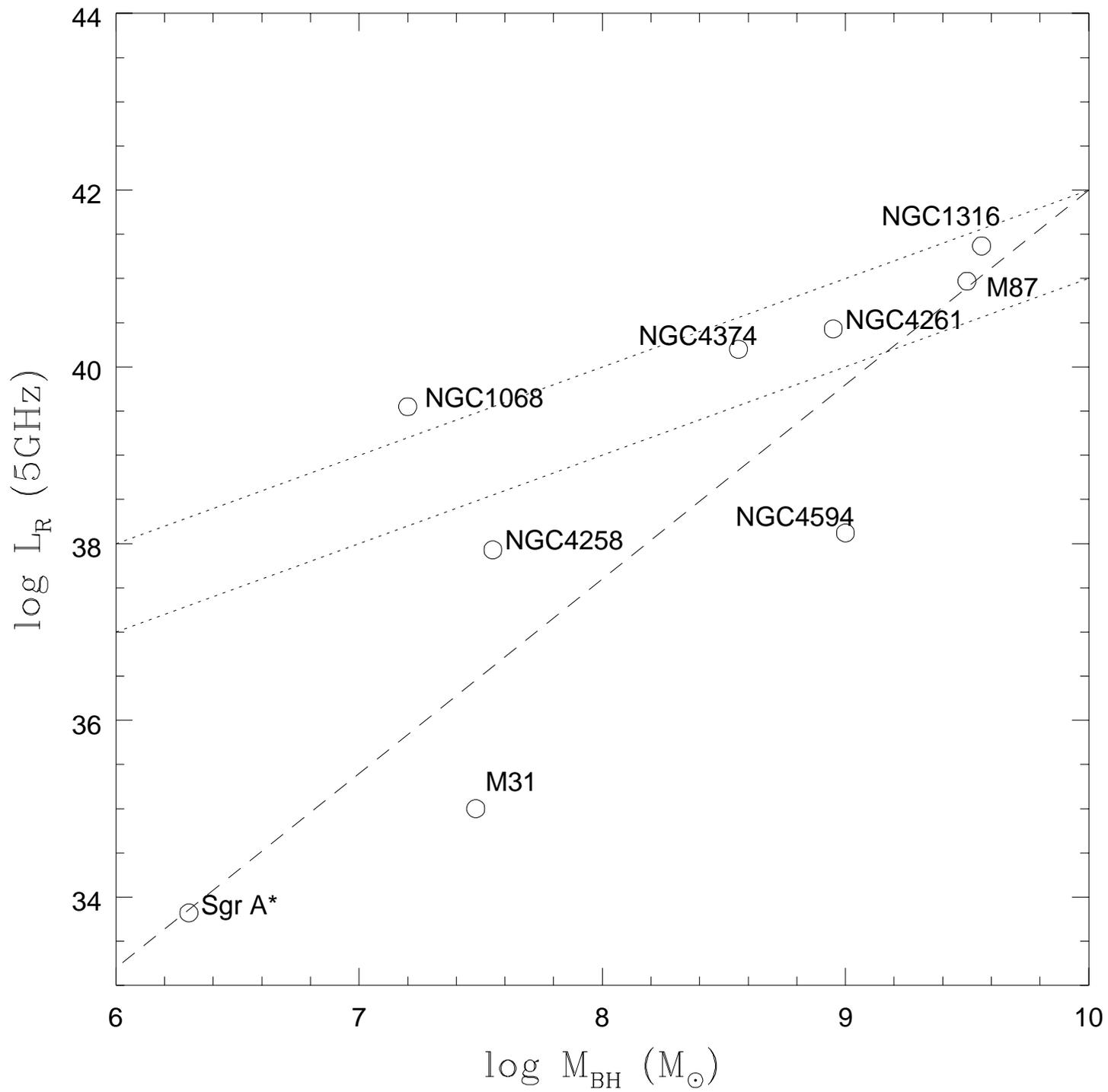
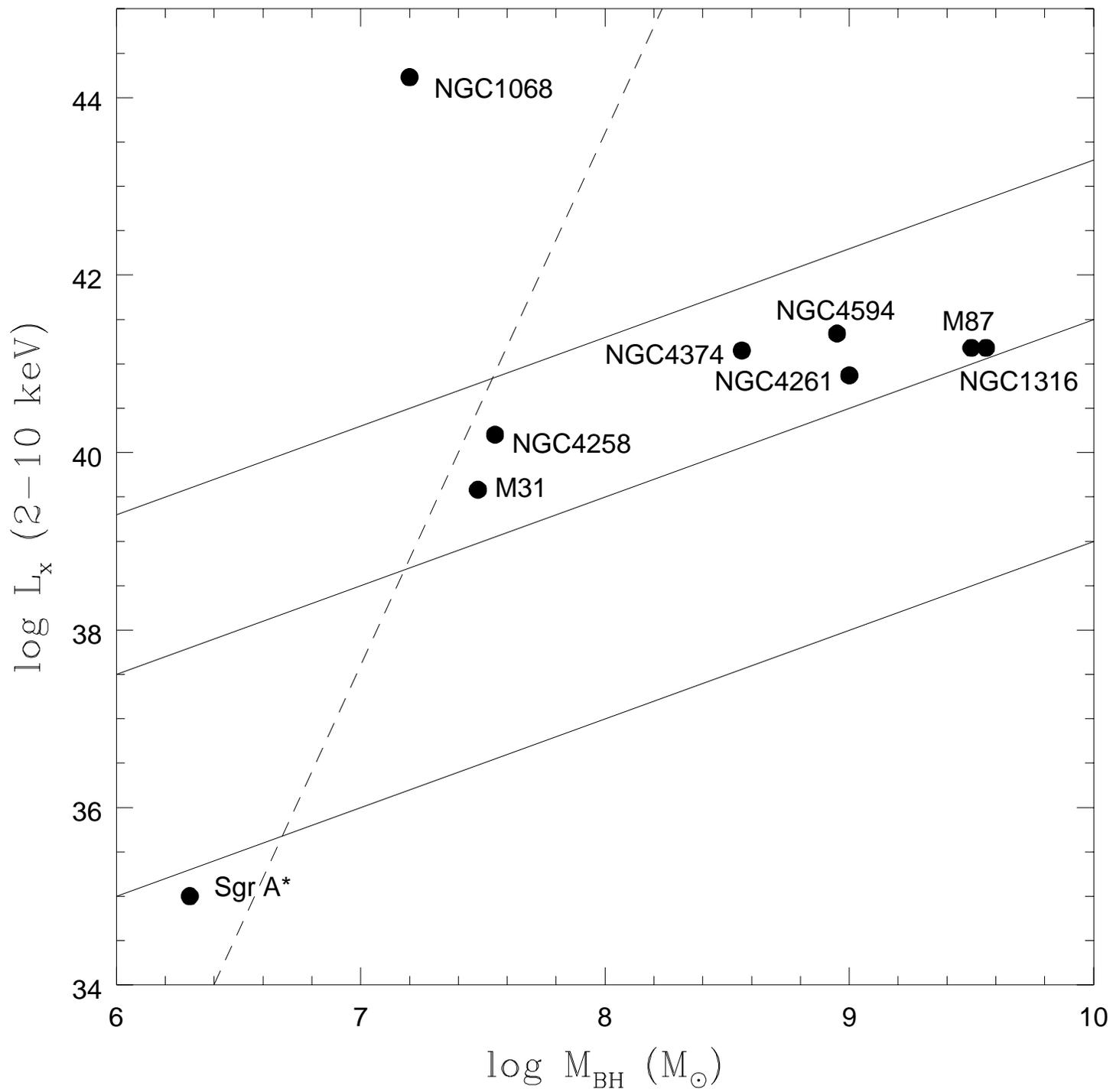
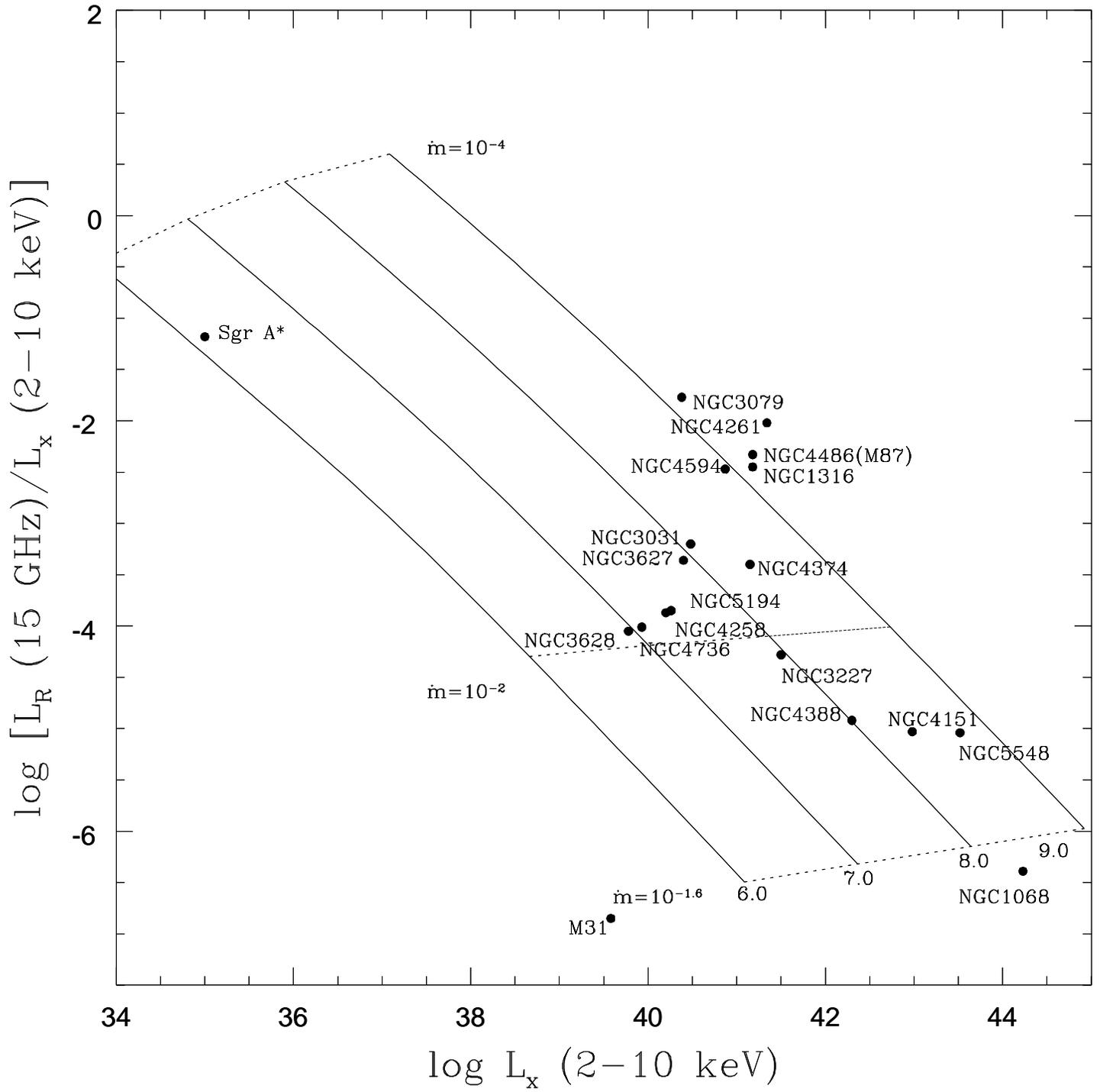
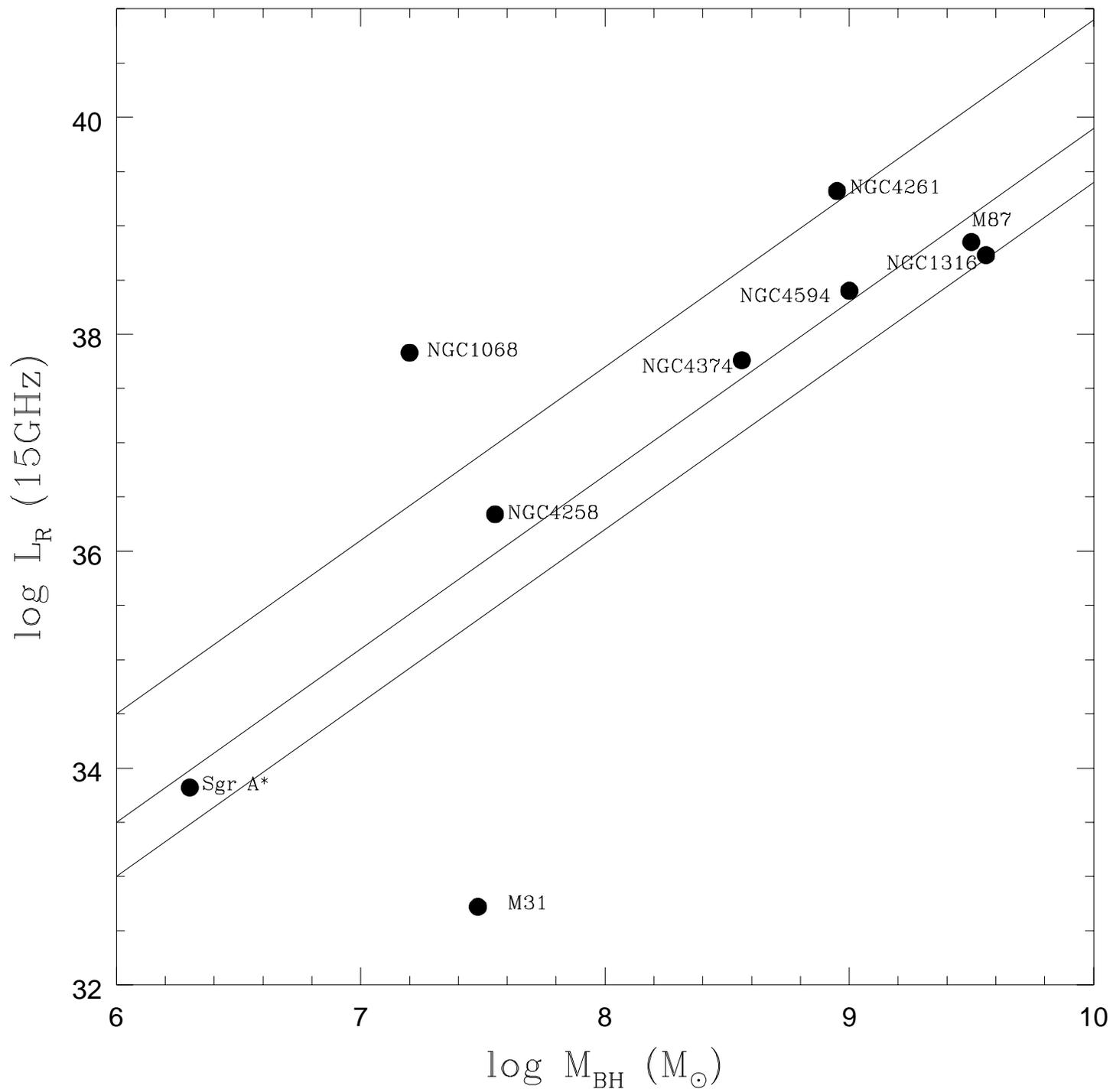
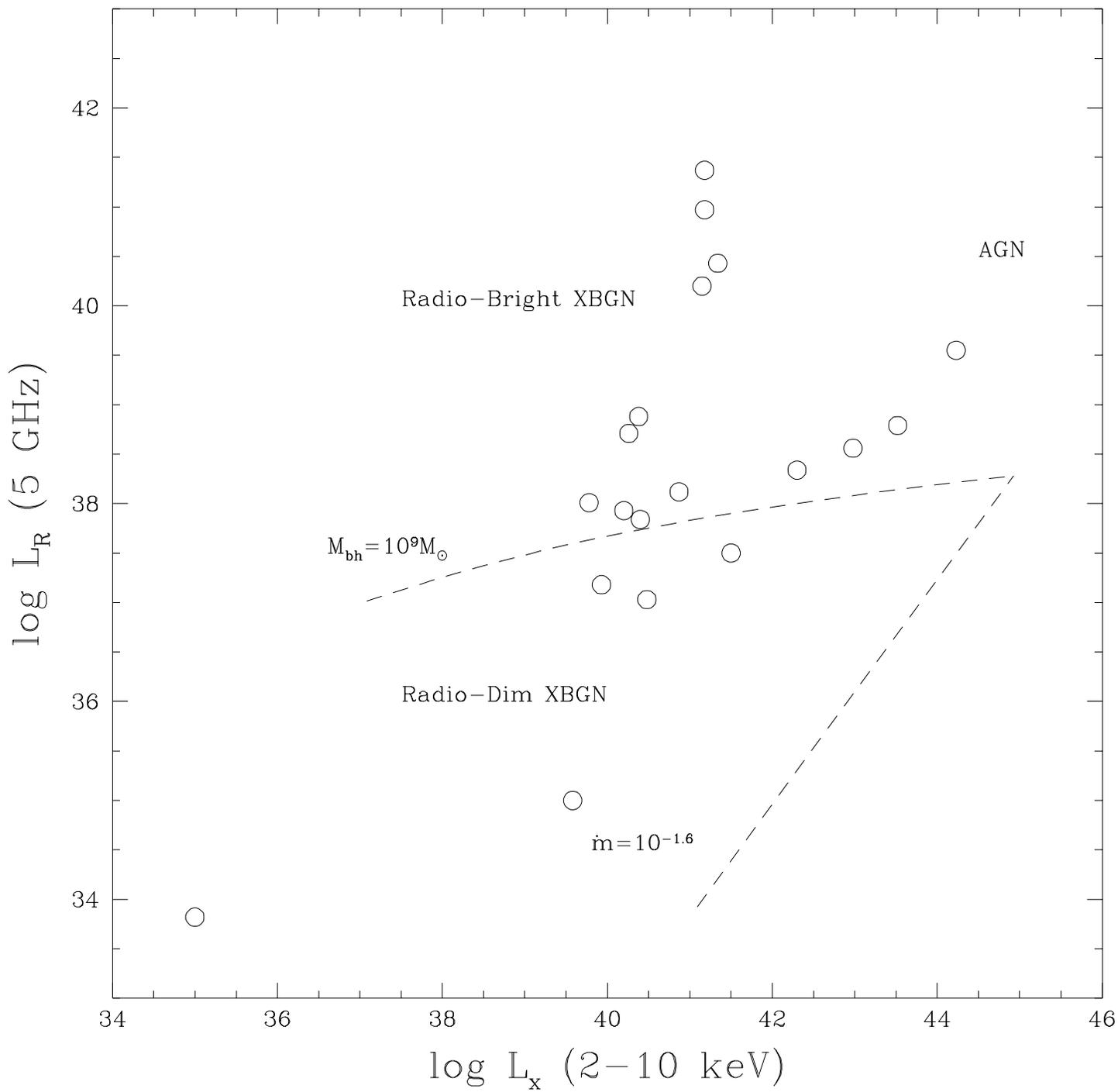
\end{document}